\documentclass{pasj00}
\draft

\begin{document}
\SetRunningHead{S. Kato}
               {Excitation of Tilt Mode on Disks with Eccentric and Misaligfned Secondary}
\Received{2014/0/00}%{yyyy/mm/dd}
\Accepted{2014/0/00}%{yyyy/mm/dd}

\title{Simultaneous Resonant Excitation of Low-frequency Eccentric Wave and Tilt Wave on Tidally Deformed Disks} 

%%% begin:list of authors
\author{Shoji \textsc{Kato}}%
%\thanks{Example: Present Address is xxxxxxxxxx}
%  }
\affil{2-2-2 Shikanodai-Nishi, Ikoma-shi, Nara 630-0114}
%  Nara 636-8503}
\email{kato.shoji@gmail.com; kato@kusastro.kyoto-u.ac.jp}

%\email{ccccc@xxx.xxx.xx.xx}
%%% end:list of authors

%%% Please use the following style in case that sorting by 
%%% affilation is impossible. 
%
% \author{%
%   D-Firstname \textsc{D-Familyname}\altaffilmark{1}
%   E-Firstname \textsc{E-Familyname}\altaffilmark{1,2}
%   and
%   F-Firstname \textsc{F-Familyname}\altaffilmark{2}}
% \altaffiltext{1}{Address of Institute}
% \email{ddddd@xxx.xxx.xx.xx}
% \email{eeeee@xxx.xxx.xx.xx}
% \altaffiltext{2}{Address of Institute}

%% `\KeyWords{}' always has to be placed before `\maketitle'.
\KeyWords{accretion, accretion disks --- Be/X-ray binaries --- instabilities
    --- waves} 
%Do NOT move this preamble from here!

\maketitle

\begin{abstract}
Simultaneous excitation of low-frequency eccentric precessing mode (one-armed p-mode) and tilt mode 
on tidally deformed disks is considered.
If the orbit of the secondary star is eccentric and its orbital plane is misaligned with the disk plane of the primary,
the above-mentioned two low-frequency oscillation modes are simultaneously excited on the primary disk, the former having prograte precession and the latter having retrograde precession.
This excitation of disk oscillations is due to a wave-wave resonant excitation process considered by Kato (2013).
If parameter values relevant to Be/X-ray binary systems are adopted, the periods of these excited oscillations are around ten times 
of the orbital period of the secondary, which may be comparable with the time scale of giant outbursts
observed in Be/X-ray systems. 
\end{abstract}

%\noindent IMPORTANT NOTICE\\
%1. ``\verb|\draft|'' creates single column and double spaces format.\\
%2. If you comment out ``\verb|\draft|'', the output will be double column
%   and single space.\\
%3. For cross-references, the use of ``\verb|\label|, \verb|\ref|, 
%\verb|\cite|'' 
%   and the thebibliography environment is strongly recommended. \\
%4. Do NOT use ``\verb|\def|, \verb|\renewcommand|''.\\
%5. Do NOT redifine commands provided by PASJ00.cls.\\

\section{Introduction}

Be/X-ray binaries consist of a Be star (primary) and a compact object (a neutron star in general, secondary)(Reig 2011).
In some Be/X-ray binaries  the secondary is known to  have an eccentric ($e\not= 0$) 
orbit whose rotational axis is misaligned ($\delta\not= 0$) with the spin axis of the primary.
Such a situation wil be naturally expected, since supernovae explosions leading to Be/X-ray 
binaries will not be always  symmetric. 
Be stars are rapidly rotating close to their break-up velocity and decretion disks are formed in their
equatorial plane.    
Since the axis of decretion disks are misaligned with the axis of the orbital plane, 
the decretion disks are warped in thier outer region so that the disk planes tend to 
the orbital plane of the secondary.
In the inner part of the disks, however, the disk plane is still close to the equatorial plane of the Be stars, 
since the viscous timescale is shorter than the tidal timescale there (Martin et al. 2011).

Various types of long-term variations have been observed in Be/X-ray binaries. 
For example, two kinds of quasi-periodic outbursts, i.e., normal (type I) outbursts and
giant (type II) outbursts.
The normal outbursts occur around periastron passage of the compact object, and are considered to come from enhancement 
of mass transfer from the Be disk to the compact secondary  (Okazaki \& Neguerule 2001; Neguerule \& Okazaki 2001).
The giant outbursts seem to be less understood compared with the normal outbursts.
However, many authors suggest that they are due to mass transfer from the Be disk to the compact object by the latter 
passing through inside the Be disk which is misaligned and warped (Martin et al. 2011; Okazaki et al. 2013;
Moritani et al. 2013).

The observed secular change from Be star to Be shell star and vice-versa are also supposed to be a result of 
variations of tilt angle of misaligned disks (Martin et al. 2011).
Even in a single Be star, so-called V/R spectrum variations have been observed, 
which are considered to be due to
one-armed eccentric precessing waves in disks (Okazaki 1991; see also Kato 1983).

The purpose of this paper is to deepen our understanding of the causes of long-term variations in  
Be/X-ray binaries by suggesting one of possible mechanisms of  long-term variations of the disks.
We have already suggested that long-term variations can be excited on tidally deformed disks.
That is, the eccentric precession mode (one-armed nearly horizontal p-mode oscillation) and
the tilt mode (one-armed vertical p-mode oscillation), both being global and low-frequency oscillation modes, can be excited 
on tidally deformed disks by a wave-wave resonant process described by Kato (2013b) (see also Kato et al. 2011, Kato 2004, 2008).\footnote{
The wave-wave coupling process can be regarded as an extension of the mode-mode coupling process by Lubow (1991).
}
Excitationon of the former mode (the eccentric prcession mode) comes from a resonant coupling between the mode and a disk deformation
through a high-frequency disk-oscillation mode and 
is considered to be the cause of superhumps observed in dwarf novae (Lubow 1991, Kato 2013a).
Excitation of the latter mode (the tilt mode) also comes from a resonant coupling between the mode and a disk deformation through 
another  high-frequency disk-oscillation mode,
and is regarded as one of posible causes of negative superhumps observed in dwarf novae (Lubow 1992, Kato 2014b).
Excitation of these global and low-frequency oscillation modes by the processes mentioned above are also expected in disks of  
Be/X-ray binaries, and further studies along this line will be worthwhile.
The purpose of this paper, however, is to suggest an another coupling process which can simultaneously excite both low-frequency oscillation modes; 
a direct coupling between the two low-frequency modes through a disk deformation.
This simultaneous excitation of the two low-frequency oscillation modes  does not always occur.
A particular orbit configuration of the secondary is required, but such configuration is expected in the cases of Be/X-ray binaries.

The requirement is that the disks are subject to a secondary star whose orbit is eccentric and whose orbital axis is misaligned with
the spin axis of the primary.
That is, although the tidal force brings about various time- and azimuthal-dependent deformations on disks, 
the disks subject to the above-mentioned secondaries have a two-armed deformation, 
even when time-averaged disks are considered (see appendix 1).
In other words,  the time-averaged disk is not axisymmetric.
Through this two-armed pattern of deformation, the set of two low-frequency oscillation modes (eccentric precession mode and tilt mode)    
are excited simultaneously.
In sections 2 and 3, the resonant conditions, and frequencies of oscillations resulting from the resonant coupling are
discussed, respectively.
In section 4, numerical results are presented, and section 5 is devoted to discussions.

\section{Two Low-Frequency Oscillation Modes and Their Resonant Coupling}

Let us consider axisymmetric geometrically thin disks, no radial flow being considered.
On such disks, a small amplitude oscillation is superposed whose frequency is $\omega$ and the azimuthal wavenumber
is $m$.
The displacement vector, 
associated with the oscillation
is written in the form 
\begin{equation}
       {\mbox{\boldmath $\xi$}}(\mbox{\boldmath $r$},t)=\Re\biggr[\breve{\mbox{\boldmath $\xi$}}(r,z){\rm exp}[i(\omega t-m\varphi)]\biggr],
\label{2.1}
\end{equation}
where $\Re$ denotes the real part.
Here, $\mbox{\boldmath $r$}$ is the cylindrical coordinates ($r$, $\varphi$, $z$), whose center is at the disk center
(this is also the center of the primary star) and the $z$-axis is the rotation axis of the disk.

\subsection{Eccentric precession modes (One-armed low-frequency p-mode oscillations)}

We consider oscillations whose frequency is $\omega$ and whose motions are mainly parallel to the equatorial plane 
(i.e., $n=0$, see footnote 2 for the meaning of $n=0$) 
(p-mode oscillations).
Their propagation region in the radial direction is then specified by 
$(\omega-m\Omega)^2>\kappa^2$,~\footnote{
This can be seen from the local dispersion relation (e.g., Okazaki et al. 1987, Kato 2001)
$$
       [(\omega-m\Omega)^2-\kappa^2][(\omega-m\Omega)^2-n\Omega_\bot^2]=k^2c_{\rm s}^2(\omega-m\Omega)^2,
                    \nonumber
$$
where $\Omega_\bot$ is the vertical epicyclic frequency, $n$ is the vertical node number of the radial component of 
the displacement vector, $\xi_r$,
$k$ is the radial wavenumber of oscillations, and $c_{\rm s}$ is the acoustic speed in the disks.
It is noted that the vertical component of the displacement vector, $\xi_z$, has one less 
vertical node number compared with that of $\xi_r$.
}
where $\Omega(r)$ is the angular
velocity of disk rotation on the equatorial plane and $\kappa(r)$ is the radial epicyclic frequency defined by 
$\kappa^2=2\Omega(2\Omega+rd\Omega/dr)$.

In the case of one-armed ($m=1$), slowly precessing p-mode oscillations,  the propagation region is given 
by $\omega<\Omega-\kappa$.
In disks where tidal force dominates over pressure force,  we have 
$\Omega-\kappa>0$,  and the difference between $\Omega$ and $\kappa$ increases with increase of radius
[see equation (\ref{2.3})].
Hence,  the propagation region of 
the one-armed oscillation with a given frequency $\omega (>0)$ (prograde) is outside the radius specified by $\omega=(\Omega-\kappa)_{\rm c}$,  as is shown schematically in figure 1.
Here, the subscript c denotes the value at radius $r_{\rm c}$ (capture radius) where $\omega=\Omega-\kappa$.
Inside the radius $r_{\rm c}$, the oscillation spatially damps.
This mode is denoted hereafter mode E (eccentric mode) and the frequency is denoted by $\omega_{\rm E}$.
Instead of $\omega_{\rm E}$, the mode E is sometimes specified by $r_{\rm c}$, 
since $\omega_{\rm E}$ and
$r_{\rm c}$ is related by $\omega_{\rm E}=(\Omega-\kappa)_{\rm c}$. 

In order to specify oscillation modes in general, however, we need the node number in the vertical direction, $n$, 
of the radial component, $\xi_r(\mbox{\boldmath $r$},t)$, of the displacement vector,
$\mbox{\boldmath $\xi$}(\mbox{\boldmath $r$}, t)$.
In the present mode, the oscillations are mainly on the equatorial plane, and we take $n=0$.
In summary, the mode E is specified by the set of ($\omega_{\rm E}$, $m_{\rm E}$, $n_{\rm E}$) with $m_{\rm E}=1$ 
and $n_{\rm E}=0$, where $\omega_{\rm E}=(\Omega-\kappa)_{\rm c}$.
It is noticed that this oscillation mode is known to be the mode causing superhumps in dwarf novae (Osaki 1985).

\subsection{Tilt modes}

Second, we consider one-armed ($m=1$) tilt mode. 
In this mode the disk plane oscillates time-periodically up and down in the vertical direction, i.e., $n=1$.
This mode is sometimes called a corrugation wave (e.g., Kato 1989).
The propagation region of the mode in the radial direction is specified by $\omega<\Omega-\Omega_\bot$ [see from footnote 2 that $(\omega-m\Omega)^2-\Omega_\bot^2>0$ is propagation regions of oscillations when $\Omega_\bot$ is larger than $\kappa$, and see from equations (\ref{2.3}) and (\ref{2.5}) that $\Omega_\bot$ is always larger than
$\kappa$),
where $\Omega_\bot(r)$ is the vertical epicyclic frequency.
In general, since $\Omega<\Omega_\bot$ and the difference between $\Omega$ and $\Omega_\bot$ increases with increase of $r$ in tidally deformed disks [see equation (\ref{2.5})],
the tilt mode is retrograde ($\omega<0$) and its propagation region is inside the radius,
$r_{\rm t}$, specified by $\omega=(\Omega-\Omega_\bot)_{\rm t}$ as is shown in figure 1, 
where the subscript denotes the value at $r_{\rm t}$.
The radius $r_{\rm t}$ is a turning point separating the  propagation region and the evanescent region of 
oscillations.
This mode is denoted hereafter mode T (tilt).
This tilt mode is specified by a set of ($\omega_{\rm T}$, $m_{\rm T}$, $n_{\rm T}$) with $m_{\rm T}=1$ and $n_{\rm T}=1$, and  $\omega_{\rm T}=(\Omega-\Omega_\bot)_{\rm t}$.
To specify the mode,  sometimes $r_{\rm t}$ is used instead of $\omega_{\rm T}$, since they are related by
$\omega_{\rm T}=(\Omega-\Omega_\bot)_{\rm t}$. 
It is noticed that this tilt mode is supposed to be one of possible  modes causing negative superhump in dwarf novae.

\subsection{Expressions for $\kappa$ and $\Omega_\bot$}

%---------------------- Figure 1 -----------------------------------
\begin{figure}
\begin{center}
    \FigureFile(80mm,80mm){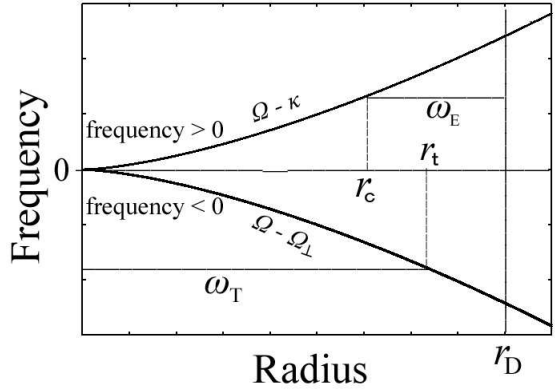}
    %%% \FigureFile(width,height){filename}
\end{center}
\vspace{40pt}
\caption{Propagation diagram schematcally showing trapped regions of $\omega_{\rm E}$ and $\omega_{\rm T}$ 
oscillations on the radius-frequency plane.
The $\omega_{\rm E}$ oscillation is trapped bwtween $r_{\rm c}$ and $¥r_{\rm D}$ (outer edge of the disk),
while the  $\omega_{\rm T}$ oscillation is  trapped inside $r_{\rm t}$.
 In the lowest order of approximations the radial distributions of $\Omega-\kappa$ and $\Omega-\Omega_\bot$
are the same wth oppsite signs.}
\end{figure}
%-------------------------------------------------------------------

Next, we should derive detailed expressions for $\kappa(r)$ and $\Omega_\bot(r)$.
The disks are subject to tidal potential, $\psi_{\rm D}(\mbox{\boldmath $r$},t)$, of the secondary.
In the cases where the orbit of the secondary is eccentric ($e\not= 0$) and the axis of the orbital plane is misaligned ($\delta\not= 0$) with
the axis of the disk plane of the primary, $\psi_{\rm D}(\mbox{\boldmath $r$},t)$ has various time- and azimuthal-dependences.
What we need to evaluate $\kappa(r)$ and $\Omega_\bot(r)$ are the time- and azimuthal-averaged part of $\psi$,
which is denoted here $\langle\psi_{\rm D}\rangle_{\varphi, t}$.
The force balance in the radial direction on the equatorial plane ($z=0$) in averaged disks is given by
$-GM/r^2-(\partial\langle\psi_{\rm D}\rangle_{\varphi, t}/\partial r)_{z=0}+\Omega^2r=0$, where
the subscript $z=0$ shows the value on the equatorial plane ($z=0$), and $\Omega$
is the angular velocity of disk rotation on the equatorial plane in time- and azimuthal-averaged disks.
The results concerning $\langle\psi_{\rm D}\rangle_{\varphi,t}$ given in appendix 1 [equation (\ref{A1.23})] 
show that $\Omega^2$ on the equatorial plane is expressed as [equation (\ref{A1.24})]
\begin{equation}
     \Omega^2=\Omega_{\rm K}^2-\frac{GM_{\rm s}}{2a^3}\biggr[\biggr(1+\frac{3}{2}e^2\biggr)+
            \frac{9}{8}(1+5e^2)\biggr(\frac{r}{a}\biggr)^2+...\biggr],
\label{2.2}
\end{equation}
where $\Omega_{\rm K}$ is the angular velocity of the Keplerian rotation, i.e., $\Omega^2_{\rm K}=GM/r^3$, 
$M_{\rm s}$ is the mass of the secondary,   and the
terms up to the order of $e^2$ and $(r/a)^2$ are taken, $a$ being the mean separation distance between the primary and secondary.
From this expression for $\Omega^2$ we can derive $\kappa^2$ defined by $\kappa^2=2\Omega(2\Omega+rd\Omega/dr)$.
Then, assuming that the tidal force is weak, we have (see appendix 1) 
\begin{equation}
      \Omega-\kappa=\frac{3}{4}q \Omega_{\rm K}\biggr(\frac{r}{a}\biggr)^3\biggr[\biggr(1+\frac{3}{2}e^2\biggr)
                  +\frac{15}{8}(1+5e^2)\biggr(\frac{r}{a}\biggr)^2\biggr],
\label{2.3}
\end{equation}
within the same approximations mentioned above, where $q$ is the mass ratio given by $q=M_{\rm s}/M$

The vertical derivative of $\langle\psi_{\rm D}\rangle_{\varphi, t}$, i.e., 
$\partial\langle\psi_{\rm D}\rangle_{\varphi, t}/\partial z$,  has a part which is proportional to $z$, 
which gives the part of $z\Omega_\bot^2$ resulting from the tidal force.
From the results in appendix 1 we have  
\begin{equation}
   \Omega_\bot^2=\Omega_{\rm K}^2+\frac{GM_{\rm s}}{a^3}\biggr[\biggr(1+\frac{3}{2}e^2\biggr)
                 +\frac{9}{4}(1+5e^2)\biggr(\frac{r}{a}\biggr)^2\biggr].
\label{2.4}
\end{equation}
Using this, we have 
\begin{equation}
       \Omega-\Omega_\bot=-\frac{3}{4}q\Omega_{\rm K}\biggr(\frac{r}{a}\biggr)^3\biggr[\biggr(1+\frac{3}{2}e^2\biggr)
                 +\frac{15}{8}(1+5e^2)\biggr(\frac{r}{a}\biggr)^2\biggr].
\label{2.5}
\end{equation}
It should be noticed that $\Omega-\kappa$ and $\Omega-\Omega_\bot$ have the same expressions with the opposite signs in the present order of approximations.

\section{Resonant Coupling of Two Low-Frequency Oscillations and Their Excitation}

In section 2, we have discussed two low-frequency modes of oscillations.
In this section we consider first in what cases they can resonantly interact with each other through the disk deformation,
and then examine whether the resonance can excite the oscillations.

\subsection{Resonant conditions}

In general, three oscillation modes with frequencies $\omega_{\rm i}$ (where ${\rm i}=1,2,3$) and azimuthal wavenumber 
$m_{\rm i}$ (where ${\rm i}=1,2,3$) can have a resonant interaction when $\omega_1+\omega_2+\omega_3=0$ and 
$m_1+m_2+m_3=0$ are realized. 
In the present problem, we are using the subscripts E and  T for the first two oscillations, instead of 1 and 2, rspectively.
As the third oscillation we consider disk deformation, and thus we use the subscript D instead of 3.
We consider the case of  $\omega_{\rm E}+\omega_{\rm T}=0$.\footnote{
In what disks $\omega_{\rm E}+\omega_{\rm T}=0$ is realized is examined in section 4 by estimating the
frequencies of trapped oscillations.
}
Then, $\omega_{\rm D}$ needs to be zero, 
i.e., $\omega_{\rm D}=0$, for a resonance to occur.
Furthermore, since $m_{\rm E}=1$ and $m_{\rm T}=1$, $m_{\rm D}$ need to be $-2$, i.e., $m_{\rm D}=-2$
for a resonance to occur.
In addition, the eccentric mode with frequency $\omega_{\rm E}$ is plane symmetric with respect to the equatorial plane, i.e., $n_{\rm E}=0$, 
while the tilt mode with $\omega_{\rm T}$ is asymmetric with respect to the equatorial plane, i.e., $n_{\rm T}=1$.
Nonlinear coupling between the $\omega_{\rm E}$ and $\omega_{\rm T}$ modes thus brings about an oscillation which is
asymmetric with respect to the equatorial plane.
Hence, for a coupling between the $\omega_{\rm E}$ and $\omega_{\rm T}$ modes to occur, the motions induced by tidal force must have an asymmetric part with respect to the equatorial plane,
i.e., $n_{\rm D}=1$.
For the motions to have such asymmetric part, the orbital plane of the secondary must be declined to the equatorial plane of disks.
In summary, for a resonance to occur, the disk deformation resulting from the tidal force must have $\omega_{\rm D}=0$, 
$m_{\rm D}=-2$, and $n_{\rm D}=1$.
In other words, the issue to be addressed here is whether there is a two-armed deformation in time-averaged disks when
the disk plane and the orbital plane are inclined.

Detailed calculations in appendix 1 show that this is really the case.
That is, the time-averaged tidal potential, $\langle\psi_{\rm D}\rangle_t$,
has a term [see equation (\ref{A1.30})]:
\begin{equation}
    \frac{GM_{\rm s}}{a}\biggr(\frac{r}{a}\biggr)^3\frac{15}{8}e\ \delta\ \frac{z}{r}{\rm sin}(\phi_{\rm A}-2\varphi),
\label{3.1}
\end{equation}
where $\delta$ is the inclination angle of the orbital plane of the secondary to the disk plane, $\phi_{\rm A}$ the angular distance of the periastron from
the nodal point, ${\rm N}$, along the orbit, and $\varphi$  the azimuthal coordinate of the observational point, ${\rm P}$, from the nodal point,
${\rm N}$ (see figure 5).  
%In this case, the radial component of $\mbox{\boldmath $\xi$}_{\rm D}$ has a term proportioinal to $z/r$.
This equation shows that in the cases where the axes of the orbital and disk planes are misaligned (i.e., $\delta\not= 0$)
and the orbit is eccentric (i.e., $e\not= 0$), the time-averaged disk is not axisymmetric but deformed in an two-armed form [see the term of ${\rm sin}(\phi_{\rm A}-2\varphi)$ in equation (\ref{3.1})], and 
the radial component of $\mbox{\boldmath $\xi$}_{\rm D}$ has a term proportioinal to $z/r$ [see the term proportional to $z/r$ in equation (\ref{3.1})].
The latter means the presence of $n_{\rm T}=1$.

\subsection{Excitation condition}

We have shown that the eccentric mode ($\omega_{\rm E}$ oscillation) and the tilt mode ($\omega_{\rm T}$ oscillation)
can have resonant interaction in the tidally deformed disks when the axes of disk plane and the orbital plane are
misaligned and the orbit of the secondary is eccentric.
Next, we should examine whether the resonance can excite the oscillations.
The excitation condition has been examined in a general form by Kato (2013) 
(see also Kato et al. 2011 and preceeding work by Kato 2004, 2008).
The results show that the excitation condition has a simple form of  
$(E_{\rm E}/\omega_{\rm E})(E_{\rm T}/\omega_{\rm T})>0$, where $E_{\rm E}$ and $E_{\rm T}$ are, respectively,
the wave energies of $\omega_{\rm E}$ and $\omega_{\rm T}$ oscillations, respectively.

A general expression for wave energy is given, for example, by Kato (2001, 2014a).
In the case of oscillations in geometrically thin non-magnetized disks, the sign of the wave energy is the same with the sign
of a radial average of $\omega(\omega-m\Omega)$ in the radial region where the waves are trapped (see, for example, Kato 2001).
Hence, the sign of $E/\omega$ is the same as the sign of $\omega-m\Omega$ in the region where 
the oscillations exist predominantly.
In the present problem, $\omega_{\rm E}-m_{\rm E}\Omega<0$, since the wave predominantly exists in the region where 
$\omega-m\Omega<-\kappa$ (see the previous section and figure 1).
Similarly, $\omega_{\rm T}-m_{\rm T}\Omega<0$, since the $\omega_{\rm T}$ oscillation is trapped in the region where
$\omega-\Omega<-\Omega_\bot$ (see figure 1).
Hence, the excitation condition, $(E_{\rm E}/\omega_{\rm E})/(E_{\rm T}/\omega_{\rm T})>0$, is satisfied    for 
the $\omega_{\rm E}$ and $\omega_{\rm T}$ oscillations under consideration.
 
\section{A Rough Estimate of Frequencies of Trapped Oscillations}

We have shown that the eccentric precession mode ($\omega_{\rm E}$ mode) and the tilt mode ($\omega_{\rm T}$ mode) can have resonant coupling and satisfy the resonant excitation conditions.
A remaining issue is whether the condition concerning frequencies (i.e., $\omega_{\rm E}+\omega_{\rm T}\sim 0$ or
$r_{\rm c}\sim r_{\rm t}$ ) is really realized within the disks.
To study this problem, we must examine where the $\omega_{\rm E}$ and $\omega_{\rm T}$ oscillations 
are trapped and how $\omega_{\rm E}$ (or $r_{\rm E}$) and $\omega_{\rm T}$ (or $r_{\rm T}$) depend on
parameters describing disk structure.
Then, we can see in what cases $r_{\rm c} = r_{\rm t}$ is realized.

%---------------------- Figure 2 -----------------------------------
\begin{figure}
\begin{center}
    \FigureFile(80mm,80mm){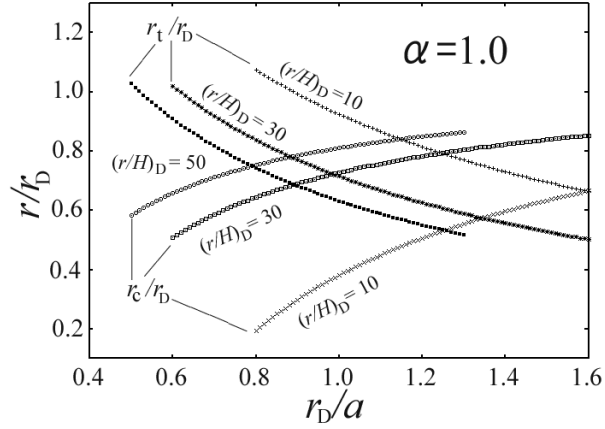}
    %%% \FigureFile(width,height){filename}
\end{center}
%\vspace{40pt}
\caption{Dependences of $r_{\rm t}$ (outer edge of trapped radius of $\omega_{\rm T}$   oscillation)   
and $r_{\rm c}$ (inner edge of trapped radius of $\omega_{\rm E}$ oscillation) on disk radius $r_{\rm D}$.
Three cases of disk thickness, i.e., $(r/H)_{\rm D}=$10, 30, and 50 are shown with $\alpha=1.0$, where
$\alpha$ specifies the radial dependence of $H$ as $H   \propto  r^\alpha$. 
}
\end{figure}
%-------------------------------------------------------------------

%---------------------- Figure 3 -----------------------------------
\begin{figure}
\begin{center}
    \FigureFile(80mm,80mm){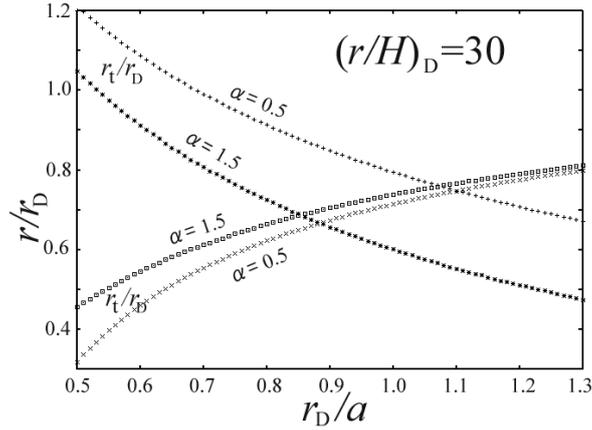} 
    %%% \FigureFile(width,height){filename}
\end{center}
%\vspace{40pt}
\caption{Dependences of $r_{\rm t}$ (outer edge of trapped radius of $\omega_{\rm T}$   oscillation)   
and $r_{\rm c}$ (inner edge of trapped radius of $\omega_{\rm E}$ oscillation) on disk radius $r_{\rm D}$.
Three cases of $\alpha$, i.e., $\alpha=$0.5, 1.0, and 1.5,  are shown with $(r/H)_{\rm D}=$ 30.
}
\end{figure}
%-------------------------------------------------------------------

First, we examine how $\omega_{\rm E}$ and $\omega_{\rm T}$ depend on parameters describing disk structure.
To do so we need to solve wave equations both for the eccentric precession mode and for the tilt mode.
We are satisfied here only with qualitative arguments, since the purpose of this paper is not to carefully solve the equations.
Hence, we solve the equations by the WKB method (see e.g., Morse \& Feshbach 1953 for the WKB method).
Analyses in appendix 2 show that the frequency, $\omega_{\rm T}$, of the tilt mode ($m=1$ and $n=1$) which is trapped 
between $r_{\rm i}\sim 0$ and 
$r_{\rm t}$ is obtained by solving  
\begin{equation}
         \int_0^{r_t} \frac{2}{c_{\rm s}}(-\omega_{\rm T}+\Omega-\Omega_\bot^2)^{1/2}(\omega_{\rm T}-\Omega+\kappa)^{1/2}dr
                =\pi\biggr(n_r+\frac{3}{4}\biggr),
\label{3.2}
\end{equation}
where $n_r$ is the node number in the radial direction, and we take here $n_r=0$.
In writing equation (\ref{3.2}) we have adopted $(\omega_{\rm T}-\Omega)^2-\Omega_\bot^2
\sim 2\Omega(-\omega_{\rm T}+\Omega-\Omega_\bot)$ and 
$(\omega_{\rm T}-\Omega)^2-\kappa^2\sim 2\Omega(-\omega_{\rm T}+\Omega-\kappa)$.
In calculating the integration in equation (\ref{3.2}), we use $c_{\rm s}=\Omega_\bot H$,    
and the half disk thickness, $H$, is taken to vary with radius as $H \propto r^\alpha$ ($\alpha$ being a constant).
Then, since $\omega_{\rm T}$ is given by $\omega_{\rm T}=(\Omega-\Omega_\bot)_{\rm t}$ and
$\Omega-\Omega_\bot$ is by equation (\ref{2.5}), equation (\ref{3.2}) can be regarded as a relation between
$r_{\rm t}/r_{\rm D}$ and $r_{\rm D}/a$ with  parameters, $e$ (eccentricity), $q(=M_{\rm s}/M)$,  
$(r/H)_{\rm D}$, and $\alpha$, where $M_{\rm s}$ is the mass of the secondary, $r_{\rm D}$ is the disk size 
and $(r/H)_{\rm D}$ denotes  the value of $r/H$ at $r_{\rm D}$. 
In this paper, as a typical example, we take, throughout the paper,  $e=0.3$ and $q=0.1$, and examine how  
the $r_{\rm t}/r_{\rm D}$ -- $r_{\rm D}/a$ relation depends on $\alpha$ and $(r/H)_{\rm D}$.
Results are shown in figure 2 for three cases of $(r/H)_{\rm D}=$10, 30 and 50 with $\alpha=1.0$, and
in figure 3 for two cases of $\alpha=0.5$, and 1.5 with $(r/H)_{\rm D}=30$.
It is noted that the value of $\omega_{\rm T}$ is obtained from $r_{\rm t}/r_{\rm D}$,  since
$\omega_{\rm T}=(\Omega-\Omega_\bot)_{\rm t}$.

In the case of one-armed eccentric precession mode ($m=1$ and $n=0$), we solve numerically the equation of the trapping condition
[see equation (\ref{WE})]:
\begin{equation}
         \int_{r_{\rm c}}^{r_{\rm D}} \frac{2^{1/2}}{c_{\rm s}}\Omega^{1/2}(-\omega_{\rm E}+\Omega-\kappa)^{1/2}dr 
                =\pi\biggr(n_r+\frac{1}{2}\biggr),
\label{3.3}
\end{equation}  
where we have adopted $(\omega_{\rm E}-\Omega)^2-\kappa^2\sim 2\Omega(-\omega_{\rm E}+\Omega-\kappa)$.
As in the case of tilt mode we adopt $e=0.3$ and $q=0.1$, and calculate $r_{\rm c}/r_{\rm D}$ as functions of $r_{\rm D}/a$ 
for some values of $(r/H)_{\rm D}$ and $\alpha$.
Three cases of $(r/H)_{\rm D}=10$, 30 and 50 with $\alpha=1.0$ are shown as a part of figure 2,
and two cases of $\alpha=0.5$ and 1.5 with $(r/H)_{\rm D}=30$ are also shown in figure 3 as a part of the figure.

Since $H=c_{\rm s}/\Omega_\bot$ and $\Omega_\bot\sim \Omega_{\rm K}=(GM/r^3)^{1/2}$, we have
$r/H\sim (GM/R_*)^{1/2}(r/R_*)^{-1/2}/c_{\rm s}$, where $R_*$ is the radius of the primary.
If we take $M=15M_\odot$, $R_*=10R_\odot$ and the disk temperature is written as $T(r)$, we have
$r/H=59\times (r/R_*)^{-1/2}(T/10^4K)^{-1/2}$.
Results of numerical simulations of Be-star disks by Carciofi \& Bjorkman (2006) show that  in the inner region of a typical Be-star disk,
$T(r)$ decreases outwards as $T\propto 1/r$ and reaches $6-8\times 10^3$ K around the radius of a few times of $R_*$.
Then, it increases outwards and outside of $\sim 10R_*$, it remains to be roughly constant of $10^4$ K.
These results show that in the inner region of disks, $r/H\propto r^0$ (i.e., $\alpha=1.0$) and $r/H\propto 1/r^{1/2}$ (i.e., $\alpha=1.5$) in the outer region.
Furthermore, if we take $r=3R_*$ and $T=7\times 10^3$ K, we have $r/H\sim 41$, and for $r=10R_*$ and $T=10^4$ K we have $r/H\sim 19$.
In addition, Martin et al. (2011) mention that observations (Wood et al. 1997) give  $(r/H)_{\rm D}\sim 25$.
These considerations suggests that the relevant ranges of $\alpha$ and $(r/H)_{\rm D}$ in Be-star disks will be $\alpha=0.5\sim 1.5$ and
$(H/r)_{\rm D} =10\sim 50$.

The case of $r_{\rm t}= r_{\rm c}$ (i.e., the case of $\omega_{\rm T}=-\omega_{\rm E}$)
is particularly of interest.
If $(r/H)_{\rm D}$ and $\alpha$ are fixed,  $r_{\rm t}=r_{\rm c}$ is realized when the disk has a particular radius,
as we can see from figures 2 and 3.
The relation between the period of $\omega_{\rm E}$ oscillation (which is the same with the period of $\omega_{\rm T}$ oscillation 
with the oppposite sign) 
and the disk radus, $r_{\rm D}/a$,  in the cases of $r_{\rm c}=r_{\rm t}$  are shown in figure 4 for three cases of 
$\alpha=0.5$, 1.0, and 1.5  by changing the value of $(r/H)_{\rm D}$.
The value of $(r/H)_{\rm D}$  on the curves are shown by attaching labels.
The periods are normalized by the orbital period of the secondary in the observational frame, $2\pi/\Omega_{\rm orb}$,
where $\Omega_{\rm orb}=(1+q)^{1/2}(GM/a^3)^{1/2}$.
We have adopted here period instead of frequency,  since period will be better than frequency for comparison with
observations.

%---------------------- Figure 4 -----------------------------------
\begin{figure}
\begin{center}
    \FigureFile(80mm,80mm){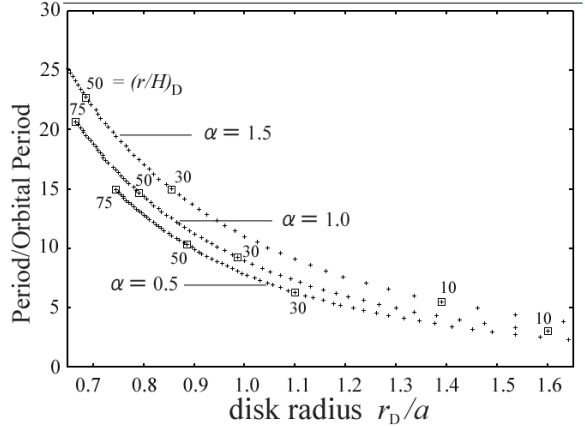}
    %%% \FigureFile(width,height){filename}
\end{center}
%\vspace{40pt}
\caption{The oscillation period   -- disk radius relation in the cases where $r_{\rm t}=r_{\rm c}$ is realized. 
The relation is shown for three cases where $\alpha$ is fixed to 0.5, 1.0, and 1.5, and    
$(r/H)_{\rm D}$ is changed. 
The value of $(r/H)_{\rm D}$ is shown by attaching on each curve.
}
\end{figure}
%-------------------------------------------------------------------

Figure 4 shows that if the disk has $\alpha=1.5$ and $(r/H)_{\rm D}=50$, for example, $r_{\rm t}=r_{\rm c}$
 (i.e., $\omega_{\rm E}=-\omega_{\rm T}$)
is realized at $r_{\rm D}/a\sim 0.7$ and the period of the oscillations is  $\sim 22$ times the orbital period.
If the disk has $\alpha=1.0$ and $(r/H)_{\rm D}=30$,  then we have 
$r_{\rm D}/a\sim 1.0$  and the period of oscillations is around 10 times the orbital period.
As an another example, if we consider a disk with $\alpha=1.5$ and $(r/H)_{\rm D}=10$, 
the resonant excitation of oscillations appears when $r_{\rm D}/a\sim 1.4$ and its frequency is about 5 times the orbital period.
In general, $(r/H)_{\rm D}$  and $\alpha$ are smaller, then $r_{\rm D}/a$ for $r_{\rm t}= r_{\rm c}$ 
is larger and sometimes $r_{\rm D}$ becomes 
larger than $a$. 

It should be noted here that if we are interested only in the cases of $\omega_{\rm E}+\omega_{\rm T}+\omega_{\rm D}=0$,
the disk radius, $r_{\rm D}$, required becomes larger than $a$ for some parameter values as mentioned above  and shown in figure 4. 
Such disks with large radius may not be realized. 
However, we should remember here that in non-pressureless disks, the resonance will not be strictly restricted only when 
$\omega_{\rm E}+\omega_{\rm T}+\omega_{\rm D}=0$.
That is, even when there is a small frequency deviation from the condition of $\omega_{\rm E}+\omega_{\rm T}+\omega_{\rm D}=0$, 
the resonance occurs and oscillations will be excited, although the growth rate of oscillations may be the highest around the case of 
$\omega_{\rm E}+\omega_{\rm T}+\omega_{\rm D}=0$.
In other words, in real situations, especially in disks with high temperature,  the resonance occurs in  a finite width in frequency space and the value of $r_{\rm D}/a$ required for resonant excitation may not be so severe as discussed above.
This will be mentioned in the next section.

\section{Summary and Discussion}

Kato (2013b, see also Kato 2004, 2008, and Kato et al. 2011) showed that in deformed disks a pair of trapped oscillaions
which are resonantly coupled through disk deformation are resonantly excited if
$(E_1/\omega_1)(E_2/\omega_2)>0$.
Here,  ($\omega_1$, $E_1$) and ($\omega_2$, $E_2$) represent, respectively,  
the set of frequency and wave energy of each trapped oscillation.
This condition of resonant excitation  of trapped oscillations in deformed disks can be extended to magnetized disks (Kato 2014a).
We examined in this paper from the viewpoint of frequency of oscillations whether this wave excitation process can be considered 
to be one of possible causes of long-term variations observed in Be/X-ray binary systems.

In this paper we have considered resonant couplings of two low-frequency oscillation modes (i.e., $\omega_{\rm E}$ oscillation
mode and $\omega_{\rm T}$ oscillation mode)
and showed that they are excited simultaneously when the disks are subject to a secondary star with eccentric orbit whose orbital plane
is inclined from the disk plane.
There are some reasons why we are interested in such special cases in this paper. 
First, the cases are theoretically of interest, since oscillation modes contributing to resonance are both low-frequency ones,  different from the cases considered before (see Introduction).
Because of this difference, the resonant condition can be satisfied only when binary systems have 
special configurarions.
Second, different from the two cases considered before, excitation of oscillations occur when disk radius, $r_{\rm D}$,  is comparable with the mean orbital radius of the secondary, $a$.
%It is noted that Okazaki et al. (2013) and Moritani et al. (2013) considered that the outbursts of Be/X-ray binaries
%occur in such a situation as $r_{\rm D}\sim a$.  

The two kinds of global, low-frequency oscillation modes in disks are 
i) the one-armed p-mode oscillation ($m=1$ and $n=0$) with prograde precession in the azimuthal direction
(Osaki 1985, see also Kato 1983), and ii) the the tilt mode ($m=1$ and $n=1$) with retrograde precession.
Local frequencies of these two oscillations are the same with the opposite signs within the approximations including the terms
up to the order of $e^2$ and the second term in the expansion with respect to $r/a$ [see equations (\ref{2.3}) and
(\ref{2.5})].
Both oscillations have the same sign of $E/\omega$, and thus they are excited by resonant coupling, if
the disks have two-armed (i.e., $m_{\rm D}=-2$), steady deformation ($\omega_{\rm D}=0$).
Such deformation is really expected when the orbit of the secondary is eccentric ($e\not= 0$) and the orbital plane is 
misaligned from the disk plane ($\delta\not= 0$) (see appendix 1).

In Be/X-ray binaries, the misalignement will be expected since the disk around the Be star is a decretion one.
In the outer part of the disks, however, they will be warped towards the binary orbital plane. 
The critical radius (tidal warp radius) where the disk is warped has been estimated by comparing tidal torque and 
viscous torque by Martin et al. (2011).
They show that in the case of Be/X-ray systems with long period, the warp radius will be outside the disk.
They suggest that disks are relatively flat inside the radius, but tilted from the equator of the Be star and precess.
Although the reasons may be different, the resonant coupling processes considered in this paper support the idea that
the disks inside the tidal warp radius are tilted and precess.  

We have estimated the frequency of the excited oscillations.
As is shown in figure 4, the expected periods of such low-frequency oscillations are around ten times the
orbital periods.
In Be/X-ray systems, two kinds of outbursts have been observed (e.g., Reig 2011), i.e., normal outbursts
and giant outbursts.
The former is considered to be due to periastron passage of the secondary star (Okazaki \& Negueruela 2001; Negueruela \&
Okazaki 2001).
The cause of the latter seems to be less understoodand, but many reserachers seem to consider that it is related to  
interactions between secondary star and precessing warped or tilted disk (Martin et al. 2011; Okazaki et al. 2013: 
Moritani et al. 2013).
The period of low-frequency trapped oscillations which are resonantly excited in disks by the present resonant process seems to be comparable 
with the period of the giant outbursts in  Be/X-ray systems.  

The radius where the resonant condition $r_{\rm t}=r_{\rm c}$ (i.e., $\omega_{\rm E}+\omega_{\rm T}=0$) is realized, however,  seems to 
be large and rarther larger than $a$ in some cases, as is shown in figure 4.  
Related to this point, we should notice that the resonant excitation is not restricted only to the cases of the exact resonance of 
$\omega_{\rm E}+\omega_{\rm T}=0$ ($\omega_{\rm D}=0$ in the present problem) in high temperature disks, 
although we have focused in this paper our attention only on the cases of $\omega_{\rm E}+\omega_{\rm T}=0$.
That is, in a certain finite range of $r_{\rm D}$ where $\omega_{\rm E}+\omega_{\rm T}$ is slightly deviated from zero, the resonance occurs.
To understand this situation, let us remember a difference between pressureless disks and those with finite pressure. 
In pressureless disks, the resonance is restricted exactly to the case of
$\omega_{\rm E}+\omega_{\rm T}+\omega_{\rm D}=0$, but in disks with a finite temperature,
the resonance occurs even when $\omega_{\rm E}+\omega_{\rm T}$ is slightly deviated from $-\omega_{\rm D}$ 
(which is zero in the present problem).
That is, the resonant region is broadened from a point to a range in frequency space.

In the limiting case of pressureless disks, 
the eccentric precession mode of frequency $\omega_{\rm E}$ is localized around the radius of 
$r_{\rm c}$, where $\omega_{\rm E}-\Omega=-\kappa$ (the inner Lindblad resonance), in the sense that outside the radius 
the oscillation has very short wavelength\footnote{see the local dispersion relation of disk oscillations given in footnote 2.
}
and will be damped by the presence of viscosity, and  inside $r_{\rm c}$ the amplitude of the mode sharply decreases inwards (evanescent region).
Similarly, the tilt mode of frequency $\omega_{\rm T}$ is localized around the radius of $r_{\rm t}$, since
the outside of the radius is the evanescent region of the mode and inside the radius the mode has very short wavelength.
Hence, only when $r_{\rm c}=r_{\rm t}$, the both modes can have nonlinear spatial interaction, and the interaction leads to
resonance, since in this case $\omega_{\rm E}=-\omega_{\rm T}$ and the resonant condition,
$\omega_{\rm E}+\omega_{\rm T}+\omega_{\rm D}=0$ is satisfied with $\omega_{\rm D}=0$.
In the case  where the disk has a finite temperature, however, both $\omega_{\rm E}$ and $\omega_{\rm T}$ oscillations are not localized around $r_{\rm c}$ and $r_{\rm t}$, respectively, but  
their propagation regions are widened as schematically shown in figure 1.
Corresponding to this, 
the resonance is not restricted exactly to the case of $\omega_{\rm E}+\omega_{\rm T}+\omega_{\rm D}=0$
(cf., see Meyer-Vernet \& Sicardy for broadening of the resonant region in the cases of disks with a finite temperature).
The growth rate of oscillations by the resonance, however, will be high when $r_{\rm c}=r_{\rm t}$ is realized.
This is the reason why we restricted our attention, for simplicity, to the cases of  $r_{\rm c}=r_{\rm t}$.

Analyses of this paper are qualitative, since frequency estimate of trapped oscillations was made by the WKB approximation with use of simplified disk models.
Eccentricity of the orbits considered is also only one case of $e=0.3$.
More quantitative  examinations in realistic disk models will be worthwhile.
More importantly, in the case of $r_{\rm D}\sim a$, the higher order terms neglected in this paper in  
expressions for $\Omega-\kappa$ and $\Omega-\Omega_\bot$ 
[i.e., terms which should be in the large brackets of equations (\ref{2.3}) and (\ref{2.5}), 
and are proportional to $(r/a)^4$ and so on]
should be taken into account in calculations of frequencies of trapped oscillations.
This may introduce non-negligible modification of our results in this paper, and
will be an important issue to be examined in the future.
In deriving detailed expressions for  $\Omega-\kappa$ and $\Omega-\Omega_\bot$, 
the expansion of the tidal potential, $\psi_{\rm D}$, in terms of $r/D$ (see appendix 1) will be less proper than the
expansion using the Laplace coefficients. 

Finally, we should note that in the case where the secondary has an eccentric orbit with misaligned orbital plane, 
the time and azumuthally averaged tidal potential has a non-zero component of 
$(\partial\langle\psi_{\rm D}\rangle_{t,\varphi}\partial z)_{z=0}$ [see equation (\ref{A1.26})].
The cause of appearance of such term is not clear.
This might show that the disk should be warped in misaligned systems.

\bigskip
The author thanks A.T. Okazaki for invaluable discussions and comments on Be star disks and Yasushi Nakao for
helpful comments on computational techniques.
The author also thanks the referee for careful reading of the manuscript.

\bigskip\noindent
{\bf Appendix 1. Disk Deformation and Epicyclic Frequencies}

We consider the tidal perturbations induced at a position ${\rm P}(\mbox{\boldmath $r$})$
on the disk of the primary by a scondary of mass $M_{\rm s}$.
When the point P is at a distance $R[=(r^2+z^2)^{1/2}]$ from the center of the primary star and the secondary star's
zenith distance observed at the point P is $\vartheta$ (see figure 5), the tidal gravitational
potential, $\psi_{\rm D}(\mbox{\boldmath $r$}, t)$,  at the point P is given by
(e.g., Lamb 1924)
\begin{equation}
    \psi_{\rm D}(\mbox{\boldmath $r$},t)=-\frac{GM_{\rm s}}{(D^2-2RD{\rm cos}\vartheta+R^2)^{1/2}}
       +\frac{GM_{\rm s}}{D^2}R\ {\rm cos}\vartheta,
\label{A1.1}
\end{equation}
where $D(t)$ is the distance between the primary and secondary stars at time $t$.
The second term on the right-hand side represents the potential of a uniform field of force of the
secondary star acting on the primary star.

%---------------------- Figure  -----------------------------------
\begin{figure}
\begin{center}
    \FigureFile(80mm,80mm){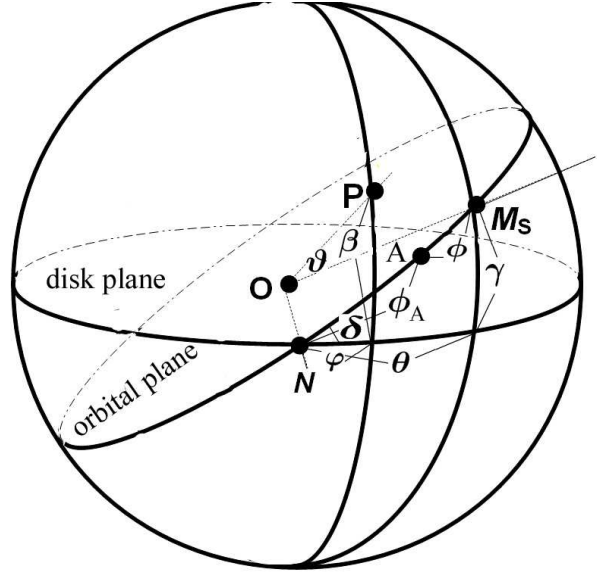}
    %%% \FigureFile(width,height){filename}
\end{center}
%\vspace{60pt}
\caption{Relation between disk plane and orbital plane of the secondary
(duplication of figure 5 of Kato 2014b) 
}
\end{figure}
%-------------------------------------------------------------------

Ristricting our attention to the cases of $R/D<1$, we expand the right-hand side of equation (\ref{A1.1}) 
by a power series of $R/D$ as\footnote{
Equation (\ref{A1.1}) can be also expanded by using the Laplace coefficients, $b_{1/2}^{(j)}$, as
\begin{equation}
    \psi_{\rm D}=-\frac{GM_{\rm s}}{2D}\sum_{j=-\infty}^{j=\infty} b_{1/2}^{(j)}(\zeta)\ {\rm cos}\ j\vartheta
         +\frac{GM_{\rm s}}{D}\frac{R}{D}{\rm cos}\ \vartheta,     
\label{A1.2}
\end{equation}
where $\zeta\equiv R/D$ and  
\begin{eqnarray}
    && \frac{1}{2}b_{1/2}^{(0)}(\zeta)=1+\frac{1}{4}\zeta^2+\frac{9}{64}\zeta^4+..., \qquad b_{1/2}^{(1)}(\zeta)=\zeta+\frac{3}{8}\zeta^3+...,
                                    \nonumber    \\
    && b_{1/2}^{(2)}(\zeta)=\frac{3}{4}\zeta^2+\frac{5}{16}\zeta^4+...,   \qquad b_{1/2}^{(3)}(\zeta)=\frac{5}{8}\zeta^3+..., \qquad
       b_{1/2}^{(4)}(\zeta)=\frac{35}{64}\zeta^4+...   \nonumber     
\label{footnote1}
\end{eqnarray}
In a previous paper (Kato 2014b), we have adopted equation (\ref{A1.1}) instead of the above expressiion (\ref{A1.2})
for $\psi_{\rm D}$.
Hence, we also adopt equation (\ref{A1.1}) in this paper.
We can obtain the same final results even if we start from equation (\ref{A1.2}).
}
\begin{equation}
   -\frac{\psi_{\rm D}}{GM_{\rm s}/D}=1+\biggr(\frac{R}{D}\biggr)^2P_2({\rm cos}\vartheta)
             +\biggr(\frac{R}{D}\biggr)^3P_3({\rm cos}\vartheta)+\biggr(\frac{R}{D}\biggr)^4P_4({\rm cos}\ \vartheta)+...,
\label{psiD2}
\end{equation}
where $P_2({\rm cos}\vartheta)$ and $P_3({\rm cos}\vartheta)$ are the Legendre polynomials
$P_\ell$ of argument ${\rm cos}\vartheta$ with $\ell=2$ and $\ell=3$, respectively.
Let us denote the polar coordinates of the point P by ($\varphi$, $\pi/2-\beta$) and the position of the secondary
star by ($\theta$, $\pi/2-\gamma$), as shown in figure 5.
It is noted that $\varphi$ is measured from the nodal point $N$.
Then, a formula of the spherical trigonometry shows that $\vartheta$ is related to them by 
\begin{equation}
   {\rm cos}\vartheta={\rm sin}\beta\ {\rm sin}\gamma+{\rm cos}\beta\ {\rm cos}\gamma\ {\cos}(\theta-\varphi).
\label{formula1}
\end{equation}

The next problem is to represent the position of the secondary star by ($\phi$, $\delta$) instead of ($\theta$, $\gamma$),
where $\phi$ is the angular distance of the secondary along the orbit, measured from the periastron, A.
In this paper we consider only the cases where the misalignement between the disk and the orbit is small, $\delta\ll \pi/2$.
Then, till the approximation of the order of $\delta^2$, we have (e.g., Kato 2014b)
\begin{equation}
      \theta=(\phi+\phi_{\rm A}), \quad {\rm sin}\gamma=\delta \ {\rm sin}(\phi+\phi_{\rm A}),
\label{delta}
\end{equation}
where $\phi_{\rm A}$ is the angular position of the periastron, A, from the nodal point, N,  along the orbit.
Then, $P_2({\rm cos}\ \vartheta)$ and $P_3({\rm cos}\ \vartheta)$ are approximated as 
\begin{eqnarray}
    P_2({\rm cos}\vartheta)\sim&&\frac{1}{4}(1-3\ {\rm sin}^2\beta)
          +\frac{3}{4}\delta\ {\rm sin}2\beta\biggr[{\rm sin}(2\phi+2\phi_{\rm A}-\varphi)+{\rm sin}\varphi\biggr]
           \nonumber \\
      +&&\frac{3}{4}{\rm cos}^2\beta\ {\rm cos}\biggr[2(\phi+\phi_{\rm A}-\varphi)\biggr],
\label{P_2approx}
\end{eqnarray}
and
\begin{eqnarray}
     P_3({\rm cos}\vartheta)\sim&& -\frac{3}{4}\delta\ {\rm sin}\ \beta\ (5\ {\rm sin}^2\beta-3)
                    \ {\rm sin}(\phi+\phi_{\rm A})
         +\frac{1}{8}(3-15\ {\rm sin}^2\beta)\ {\rm cos}\beta\ {\rm cos}(\phi+\phi_{\rm A}-\varphi)
                 \nonumber   \\
         +&&\frac{15}{8}\delta\ {\rm sin}\beta\ {\rm cos}^2\beta
           \biggr[{\rm sin}(3\phi+3\phi_{\rm A}-2\varphi)-{\rm sin}(\phi+\phi_{\rm A}-2\varphi)\biggr]
                 \nonumber  \\
         +&&\frac{5}{8}{\rm cos}^3\beta\ {\rm cos}\biggr[3(\phi+\phi_{\rm A}-\varphi)\biggr].
\label{P_3approx}
\end{eqnarray}
An expression for $P_4({\rm cos}\ \varphi)$ is omitted here, since it is somewhat lengthy.

When the orbit is eccentric, $\phi$ is not $\Omega_{\rm orb}t$, but 
\begin{equation}
         \phi=\Omega_{\rm orb}t+2e\ {\rm sin}(\Omega_{\rm orb}t)+\frac{5}{4}e^2{\rm sin}(2\Omega_{\rm orb}t)+....
\label{phi}
\end{equation}
and $D$ also varies with time as 
\begin{equation}
      \frac{a}{D}=1+ e\ {\rm cos} (\Omega_{\rm orb}t)+e^2\ {\rm cos} (2\Omega_{\rm orb}t)+...,
\label{D}
\end{equation}
where $a$ is the mean radius of the orbit and $e$ is the eccentricity of the orbit.

After these preparation, we proceed to evaluate the time and azimuthal average of $\psi_{\rm D}$
in order to calculate the horizontal and vertical
epicyclic frequencies, $\kappa(r)$ and $\Omega_\bot(r)$.
Let us denote the azimuthal average of $X$ by $\langle X\rangle_\varphi$. 
Since the $\varphi$-dependences of $\psi_{\rm D}$ come only from $P_2({\rm cos}\vartheta)$, $P_3({\rm cos}\vartheta)$,
and $P_4({\rm cos}\ \vartheta)$,..., we have
\begin{equation}
   \langle\psi_{\rm D}\rangle_\varphi=-\frac{GM_{\rm s}}{D}\biggr[1+\biggr(\frac{R}{D}\biggr)^2
                 \langle P_2({\rm cos}\vartheta)\rangle_\varphi
                 +\biggr(\frac{R}{D}\biggr)^3\langle P_3({\rm cos}\vartheta)\rangle_\varphi
                 +\biggr(\frac{R}{D}\biggr)^4\langle P_4({\rm cos}\vartheta)\rangle_\varphi+ ...\biggr],
\label{A1.20}
\end{equation}
where
\begin{eqnarray}
   &&     \langle P_2({\rm cos}\vartheta)\rangle_\varphi=\frac{1}{4}(1-3{\rm sin}^2\beta),     \\
\label{averageP2}
   &&     \langle P_3({\rm cos}\vartheta)\rangle_\varphi=\frac{3}{4}\delta\ {\rm sin}\ \beta \ (3-5{\rm sin}^2\beta)\ 
            {\rm sin}\ (\phi+\phi_{\rm A}),   \\
   &&     \langle P_4({\rm cos}\vartheta)\rangle_\varphi=\frac{3}{8}\biggr(1-5{\rm cos}^2\beta+\frac{35}{8}{\rm cos}^4\beta\biggr). 
\label{averageP3}
\end{eqnarray}

Next, we consider the time-average of $\langle\psi_{\rm D}\rangle_\varphi$, which is denoted 
$\langle\psi_{\rm D}\rangle_{\varphi, t}$.
Equations (\ref{phi}) and (\ref{D}) show that
\begin{equation}
     \biggr\langle\biggr(\frac{R}{D}\biggr)^3\biggr\rangle_t=\biggr(\frac{R}{a}\biggr)^3\biggr(1+\frac{3}{2}e^2\biggr),
\label{A1.21}
\end{equation}
\begin{equation}
     \biggr\langle\biggr(\frac{R}{D}\biggr)^4{\rm sin}(\phi+\phi_{\rm A})\biggr\rangle_t=
          \biggr(\frac{R}{a}\biggr)^4e\ {\rm sin}\ \phi_{\rm A},
\label{A1.22}
\end{equation}
\begin{equation}
      \biggr\langle\biggr(\frac{R}{D}\biggr)^5\biggr\rangle_t=\biggr(\frac{R}{a}\biggr)^5(1+5e^2),
\label{A1.22'}
\end{equation}      
where the terms till $e^2$ are taken, and $\langle X\rangle_t$ denotes the time average of $X$.
Then, we have
\begin{eqnarray}
     \langle\psi_{\rm D}\rangle_{\varphi, t}=-\frac{GM_{\rm s}}{a}
        \biggr[&&1+\frac{1}{4}(1-3{\rm sin}^2\beta)\biggr(1+\frac{3}{2}e^2\biggr)\biggr(\frac{R}{a}\biggr)^2  
                   \nonumber    \\
        &&+\frac{3}{4}e\delta(3-5{\rm sin}^2\beta){\rm sin}\beta\ {\rm sin}\ \phi_{\rm A}
              \biggr(\frac{R}{a}\biggr)^3+    \nonumber  \\
        && +\frac{3}{8}\biggr(1-5{\rm cos}^2\beta+\frac{35}{8}{\rm cos}^4\beta\biggr)(1+5e^2)\biggr(\frac{R}{a}\biggr)^4+...\biggr].
\label{A1.23}
\end{eqnarray}
This expression for $\langle\psi_{\rm D}\rangle_{\varphi, t}$ shows that
\begin{equation}
    -\biggr(\frac{\partial}{\partial r}\langle\psi_{\rm D}\rangle_{\varphi,t}\biggr)_{z=0}=
       \frac{GM_{\rm s}}{2a^3}\biggr[\biggr(1+\frac{3}{2}e^2\biggr)+\frac{9}{8}(1+5e^2)
           \biggr(\frac{r}{a}\biggr)^2\biggr]r,
\label{A1.24}
\end{equation}
where $(X)_{z=0}$ denotes the value of $X$ on the equator ($z=0$).
Hence, the force balance in the radial direction on the equator,
$-GM/r^2-(\partial\langle\psi_{\rm D}\rangle_{\varphi,t}/\partial r)_{z=0}+\Omega^2r=0$, leads to
\begin{equation}
      \Omega^2=\Omega_{\rm K}^2-\frac{GM_{\rm s}}{2a^3}\biggr[\biggr(1+\frac{3}{2}e^2\biggr)
            +\frac{9}{8}(1+5e^2)\biggr(\frac{r}{a}\biggr)^2\biggr],
\label{A1.25}
\end{equation}
where $\Omega_{\rm K}^2=GM/r^3$ and $\Omega$ is the angular velocity of disk rotation on the equatorial plane.
Taking the terms resulting from the secondary to be small, we derive $\kappa^2[\equiv 2\Omega(2\Omega+rd\Omega/dr)]$
in  the form 
\begin{equation}
     \kappa^2=\Omega_{\rm K}^2-\frac{2GM_{\rm s}}{a^3}\biggr[\biggr(1+\frac{3}{2}e^2\biggr)
             +\frac{27}{16}(1+5e^2)\biggr(\frac{r}{a}\biggr)^2\biggr].
\label{kappa}
\end{equation}
This expression for $\kappa$ leads to
\begin{equation}
      \Omega-\kappa=\frac{3}{4}q \Omega_{\rm K}\biggr(\frac{r}{a}\biggr)^3\biggr[\biggr(1+\frac{3}{2}e^2\biggr)
                  +\frac{15}{8}(1+5e^2)\biggr(\frac{r}{a}\biggr)^2\biggr],
\label{Omega-kappa}
\end{equation}
where $q=M_{\rm s}/M$.

Next, we consider $\partial\langle\psi_{\rm D}\rangle_{\varphi,t}/\partial z$.
Equation (\ref{A1.23}) gives
\begin{equation}
        -\frac{\partial}{\partial z}\langle\psi_{\rm D}\rangle_{\varphi,t}=\frac{9}{4}\frac{GM_{\rm s}}{a^2}
                      e\delta \biggr(\frac{r}{a}\biggr)^2\ {\rm sin}\ \phi_{\rm A}
         -\frac{GM_{\rm s}}{a^3}\biggr[\biggr(1+\frac{3}{2}e^2\biggr) +
                \frac{9}{4}(1+5e^2)\biggr(\frac{r}{a}\biggr)^2\biggr]z+ \O\biggr(\frac{z}{r}\biggr)^2.
\label{A1.26}
\end{equation}
The second term on the right-hand side is proportional to $z$ with a negative coefficient, bringing about a vertical harmonic oscillation of a fluid element around the equator.
In addition to this, the gravitational potential  of the primary star gives rise to a harmonic oscillation 
around the equator,
the square of the frequency being $\Omega_{\rm K}^2$.  
Hence,  the square of the vertical epicyclic frequency, $\Omega_\bot^2$,  is given by
\begin{equation}
   \Omega_\bot^2=\Omega_{\rm K}^2+\frac{GM_{\rm s}}{a^3}\biggr[\biggr(1+\frac{3}{2}e^2\biggr)
                 +\frac{9}{4}(1+5e^2)\biggr(\frac{r}{a}\biggr)^2\biggr].
\label{A1.27}
\end{equation}
This equation gives 
\begin{equation}
       \Omega-\Omega_\bot=-\frac{3}{4}q\Omega_{\rm K}\biggr(\frac{r}{a}\biggr)^3\biggr[\biggr(1+\frac{3}{2}e^2\biggr)
                 +\frac{15}{8}(1+5e^2)\biggr(\frac{r}{a}\biggr)^2\biggr].
\label{A1.28}
\end{equation}
Equations (\ref{A1.25}) and (\ref{A1.28}) show that $\Omega-\kappa$ and $\Omega-\Omega_\bot$ are the same 
till the order of $e^2$ and $(r/a)^2$, except that they have opposite signs.
In the order of $(r/a)^3$, however, a difference appears, although it is not shown here.

It should be noticed here that the first term on the right-hand side of equation ({\ref{A1.26}) is $z$-independent.
This means that in the systems where the disk plane and the orbital plane are misaligned ($\delta\not= 0$),
the disk cannot be maintained steadily in a misaligned state, unless the orbit of the secondary is circular
($e=0$) or the periastron of the orbit is just on the disk plane ($\phi_{\rm A}=0$).
In other words, this may be related to the fact that the disk must be warped.
The first term on the right-hand side of equation (\ref{A1.26}) is proportional to $(r/a)^2$, and is small in the inner region of
the disk.
Hence, a deviation from a plane symmetric disk becomes prominent in the outer part of disks 
[see Martin et al. (2011) concerning tidal warping].

The final subject to be addressed in this appendix is to 
examine whether the time-independent part of  $\psi_{\rm D}$, i.e., $\langle\psi_{\rm D}\rangle _t$,  has
a two-armed ($m_{\rm D}=\pm 2$) and asymmetric $(n_{\rm D}=1$)  part (with
respect to the equatorial plane).\footnote{
The presence of such terms has been shown in table 3 by Kato (2014b).
See the cross point of the line of $(R/a)^3$ and the column of $e^1$ in the table.
}
The presence of such terms is necessary for the twin low-frequency oscillations considered in this paper can have
resonant interaction through disk deformation.

The last term on the right-hand side of equation (\ref{P_2approx}) shows that $P_2({\rm cos} \vartheta)$ has terms 
proportinal to ${\rm sin} 2\varphi$ or ${\rm cos} 2\varphi$, but they are plane-symmetric with respect to the equator.
Hence, they are not what we are looking for.
The right-hand side of equation (\ref{P_3approx}) show that $P_3({\rm cos} \vartheta)$ has the terms which we are looking for, 
while $P_4({\rm cos} \vartheta)$ has no such terms.
Consequently, representing the terms proportional to ${\sin} 2\varphi$ or ${\rm cos} 2\varphi$ in $\psi_{\rm D}$
as
$(\psi_{\rm D})^{2\varphi}$, we have, using equations (\ref{D}) and (\ref{phi}), 
\begin{eqnarray}
       (\psi_{\rm D})^{2\varphi}=-\frac{GM_{\rm s}}{D}&&\biggr(\frac{R}{D}\biggr)^3\frac{15}{8}\delta\ {\rm sin}\ \beta\ {\rm cos}^2\beta \times
                     \nonumber \\
       \times \biggr[\biggr(&&-{\rm cos}(3\phi+3\phi_{\rm A})+{\rm cos}(\phi+\phi_{\rm A})\biggr){\rm sin}\ 2\varphi
                     \nonumber \\
         +\biggr(&& {\rm sin}(3\phi+3\phi_{\rm A})-{\rm sin}(\phi+\phi_{\rm A})\biggr){\rm cos}\ 2\varphi\biggr].
\label{A1.29}
\end{eqnarray}

The time-dependences of $D$ and $\phi$ are given by equations (\ref{D}) and (\ref{phi}), respectively.
Hence, expressing time-average of $(\psi_{\rm D})^{2\varphi}$ by $\langle(\psi_{\rm D})^{2\varphi}\rangle_t$, we have
\begin{equation}
     \langle(\psi_{\rm D})^{2\varphi}\rangle_t=
         \frac{GM_{\rm s}}{a}\biggr(\frac{r}{a}\biggr)^3\frac{15}{8}e\ \delta\ {\rm sin}\ \beta\ {\rm cos}^2\beta
                \ {\rm sin}(\phi_{\rm A}-2\varphi).
\label{A1.30}
\end{equation}
This is proportional to $z/r$ when $z/r$ is small ($n_{\rm D}=1$) (see the term of sin $\beta$).
This means that in the case of $e\not= 0$ and $\delta\not= 0$, the disk deformation has a time-independent,
two-armed ($m_{\rm D}=2$) deformation.
This deformation can have resonant interaction with two low-frequency disk oscillations considered in the text.

\bigskip\noindent
{\bf Appendix 2. Trapped Oscillations in Geometrically Thin Disks}

The velocity induced by oscillations over pure rotation (0, $r\Omega(r)$, 0) is denoted by ($u_r$, $u_\varphi$, $u_z$).
The density and pressure perturbations over the unperturbed ones, $\rho_0$ and $p_0$, are denoted by
$\rho_1$ and $p_1$, respectively.
These perturbed quantities are related by equation of motion, equation of continuity, and adiabatic relation.
By taking these perturbed quantities to be proportional to ${\rm exp}[i(\omega t-m\varphi)]$, and combining the above equations,
we have a set of equations for $u_r$ and $h_1(\equiv p_1/\rho_0$) (see, e.g., Kato et al. 2008):
\begin{equation}
     \frac{\partial h_1}{\partial r}-\frac{2m\Omega}{r(\omega-m\Omega)}h_1
      =\frac{(\omega-m\Omega)^2-\kappa^2}{i(\omega-m\Omega)}u_r,
\label{A2.1}
\end{equation}
\begin{eqnarray}
      \frac{\partial^2 h_1}{\partial z^2}&&-\frac{z}{H}\frac{\partial h_1}{\partial z}+\frac{(\omega-m\Omega)^2}{c_{\rm s}^2}h_1
               \nonumber   \\
      &&=i(\omega-m\Omega)\biggr[\frac{\partial u_r}{\partial r}+
          \biggr(\frac{\partial{\rm ln} r\rho_0}{\partial r}+\frac{m\kappa^2}{2r\Omega(\omega-m\Omega)}\biggr)u_r\biggr],
\label{A2.2}
\end{eqnarray}
where $-m^2/r^2$ has been neglected in comparison with $(\omega-m\Omega)^2/c_{\rm s}^2$, since geometrically thin disks
are considered.
Ther set of equations (\ref{A2.1}) and (\ref{A2.2}) are simultaneous partial differential equations with respect to
$h_1$ and $u_r$.
It is necessary to solve the above equations numerically in order to know detailed behaviours of trapped oscillations.
Here, however, we are satisfied by solving them approximately.

Hereafter, we assume that the disk is vertically isothermal.
Then, the partial derivative $\partial{\rm ln}(r\rho_0)/\partial r$ in equation (\ref{A2.2}) is written as 
$d{\rm ln}(r\rho_{00})/dr+(z/H)^2d{\rm ln}H/dr$, 
since in vertically isothermal disks $\rho_0(r,z)$ is stratified as $\rho_0(r,z)=\rho_{00}(r){\rm exp}(-z^2/2H^2)$, $H(r)$ being
the half-thickness of disks.
The second term with $(z/H)^2$ in the above expression for $\partial{\rm ln}(r\rho_0)/\partial r$
brings about mathematical complication in solving the above set of equations (\ref{A2.1}) and (\ref{A2.2}).
Here, we assume that the neglect of the term with $(z/H)^2$ will not bring about no essential difference in the final 
results concerning basic behaviour of trapped oscillations.
Then, the above set of equations (\ref{A2.1}) and (\ref{A2.2}) can be easily solved by reducing them to ordinary 
differential equations by a method of separation of variables.
The procedures are as follows.
First, we derive a partial differential  equation for $h_1$ by substituting equation (\ref{A2.1}) into equation (\ref{A2.2}) 
to eliminate $u_r$.
In the equation, $h_1(r,z)$ is assumed to be separable as $h_1=f(r)g(z;r)$, where $g$ is a function of $z$ 
and depends only weakly on $r$.
Then, by diviving the  equation for $h_1$ by $f(r)g(z;r)$, we can separate 
the equation into two parts, one of which is a function of $r$ alone and  the other part 
of which is a function of $z$ with a weak $r$-dependence.
Hence, as is often made in discoseismology (e.g., Ortega-Rodr\'{i}guez et al. 2008), 
by introducing a separation constant $K(r)$, we have 
\begin{equation}
     \frac{d^2g}{dz^2}-\frac{z}{H}\frac{dg}{dz}+Kg=0,
\label{A2.3}
\end{equation}
and
\begin{eqnarray}
   (\omega-m\Omega)&&\biggr[\frac{d}{dr}+\frac{d{\rm ln}r\rho_{00}}{dr}+\frac{m\kappa^2}{2r\Omega(\omega-m\Omega)}\biggr] 
     \biggr[\frac{(\omega-m\Omega)}{(\omega-m\Omega)^2-\kappa^2}\biggr(\frac{d}{dr}-\frac{2m\Omega}{r(\omega-m\Omega)}\biggr)\biggr]f
              \nonumber  \\
   &&+\biggr[\frac{(\omega-m\Omega)^2}{c_{\rm s}^2}-K\biggr]f=0.
\label{A2.4}
\end{eqnarray}
The boundary condition at infinity in the vertical direction ($z=\pm\infty$) requires that $g(z)$ is 
a Hermite polynomial and the separation constant $K$ is $n/H^2$.
Here, $n$ is zero or a positive integer (Okazaki et al. 1987) and represents the node number of $g$ in the vertical direction.

In the case of low-frequency oscillations in nearly Keplerian disks, 
$m\kappa^2/2r\Omega(\omega-m\Omega)$ and $2m\Omega/r(\omega-m\Omega)$ can be approximated to be, respectively, $-1/2r$ and $-2/r$.
Hence, assuming that $\rho_{00}(r)\propto r^{-\beta}$ ($\beta$ being constant),
we can reduce equation (\ref{A2.4}) to 
\begin{equation}
    (\omega-m\Omega)r^{2+\gamma}\frac{\partial}{\partial r}\biggr[\frac{1}{r^{2+\gamma}}\frac{\omega-m\Omega}
        {(\omega-m\Omega)^2-\kappa^2}\frac{\partial}{\partial r}\tilde{f}\biggr]
          +\frac{(\omega-m\Omega)^2-n\Omega_\bot^2}{c_{\rm s}^2}\tilde{f}=0,
\label{A2.5}
\end{equation}
where $\tilde{f}\equiv f/r^2$, $\gamma=\beta+1/2$, and $\Omega_\bot^2H^2=c_{\rm s}^2$ has been used.
In the above equation, $\omega$ and $m(=1)$ in some terms have been retained so that their origins can be traced.

Now, we introduce a new variable, $\tau(r)$, defined by 
\begin{equation}
      \tau(r)=\int_{r_i}^r r^{2+\gamma}\frac{(\omega-m\Omega)^2-\kappa^2}{(\omega-m\Omega)}dr,
\label{A2.6}
\end{equation}
where $r_i$ is the inner edge of the propagation region of oscillations.
Then, equation (\ref{A2.5}) is reduced to a simple form:
\begin{equation}
    \frac{d^2\tilde{f}}{d\tau^2}+Q(\tau)\tilde{f}=0,
\label{A2.7}
\end{equation}
where
\begin{equation}
     Q(\tau)=\frac{1}{(r^{2+\gamma})^2}\frac{(\omega-m\Omega)^2-n\Omega_\bot^2}{(\omega-m\Omega)^2-\kappa^2}
          \frac{1}{c_{\rm s}^2}.
\label{A2.8}
\end{equation}
It is noted that $Q$ is positive in the propagation region of oscillations.
That is, for the eccentric one-armed oscillations with $n=0$, $Q$ is positive since $(\omega-m\Omega)^2-\kappa^2>0$
in their  propgation region,
and for the tilt modes with $n=1$ we have $(\omega-m\Omega)^2-\Omega_\bot^2>0$ (which leads to $(\omega-m\Omega)^2-\kappa^2>0$)
in their propagation region.

We solve equation (\ref{A2.7}) by the WKB approximation.
In the case of tilt ($m=1$ and $n=1$), the outer edge of the propagation region is the radius where $Q$ becomes zero,
i.e., it is a turning point of $Q$.
Hence, near the outer edge of the propagation region, i.e., $r_t$ (see figure 1), 
we use the WKB solution for the region near a turning point (see Morse \& Feshbach 1953).
At the inner edge of the propagation region we use, for simplicity, $\tilde{f}=0$ as a boundary condtion.
Then, the WKB solution gives as the trapping condition:
\begin{eqnarray}
         W_{\rm T}&&\equiv \int_{\tau(r_i)}^{\tau(r_t)}Q^{1/2}d\tau=
               \int_{r_i}^{r_t} \frac{[(\omega-m\Omega)^2-n\Omega_\bot^2]^{1/2}[(\omega-m\Omega)^2-\kappa^2]^{1/2}}
                {c_{\rm s}\vert\omega-m\Omega\vert}dr 
                         \nonumber \\
                &&=\pi\biggr(n_r+\frac{3}{4}\biggr),
\label{A2.9}
\end{eqnarray}
where $n_r=0,1,2,...$ denotes the node number of $h_1$ (or $f_1(r)$) in the radial direction. 

In the case of one-armed precession mode, the inner edge of the propagation region, i.e., the capture radius, $r_{\rm c}$
(see figure 1), is not a turning point unlike the case of the tilt mode, if it is measured by $\tau$ (not $r$).
Rather, it is a point where $Q(\tau)$ becomes infinite [see equation (\ref{A2.8})].
Inside of $\tau=0$, $Q$ becomes infinity with minus sigh.
Hence, it might be better to adopt an standard WKB solution:
\begin{equation}
       h_1\sim \frac{1}{Q^{1/4}}{\rm exp}\biggr[\pm i\int Q^{1/2}d\tau\biggr].
\label{A2.11}
\end{equation}
We adopt $h_1=0$ at $\tau=0$ as the inner boundary condition and, for simplicity, $h_1=0$ at the outer edge of the disk, i.e., $r_{\rm D}$.
Then, the trapping condition may be written as 
\begin{eqnarray}
         W_E && \equiv \int_0^{\tau(r_t)}Q^{1/2}d\tau=\int_{r_{\rm c}}^{r_{\rm D}}                     
                          \frac{[(\omega-m\Omega)^2-\kappa^2]^{1/2}} {c_{\rm s}} dr
                                \nonumber    \\
            && =\pi\biggr(n_r+\frac{1}{2}\biggr).
\label{WE}
\end{eqnarray}
Here, $n_r=0,1,2,...$ is the node number in the radial direction.

\bigskip\noindent
{\bf Reference}

\leftskip=20pt
\parindent=-20pt
Carciofi, A.C. \ Bjorkman J.E. 2006, ApJ, 639\par
%Fu, W., \& Lai, D. 2009, ApJ, 690, 1386\par
%Fu, W., \& Lai, D. 2011, MNRAS, 410, 399 \par
%Ferreira, B.T. and Ogilvie, G.I. 2008, MNRAS, 386, 2297 \par
Kato, S. 1983, PASJ, 35, 249 \par
Kato, S. 1989, PASJ, 41, 745 \par
Kato, S. 2001, PASJ, 53, 1\par 
Kato, S. 2004, PASJ, 56, 905\par
Kato, S. 2008, PASJ, 60, 111 \par
%Kato, S. 2008b, PASJ, 60, 1387 \par
%Kato, S. 2012a, PASJ, 64, 62 \par
%Kato, S. 2012, PASJ, 64, 139\par
Kato, S. 2013a, PASJ, 65, 56\par
Kato, S. 2013b, PASJ, 65, 75 \par
Kato, S. 2014a, PASJ, 66, in press \par
Kato, S. 2014b, PASJ, 66, in press \par
%Kato, S., Fukue, J., \& Mineshige, S. 1998, Black-Hole Accretion Disks 
%  (Kyoto: Kyoto University Press)\par
%Kato, S. and Fukue, J. 2006, PASJ, 58, 909 \par  
Kato, S., Fukue, J., \& Mineshige, S. 2008, Black-Hole Accretion Disks 
  -- Toward a New paradigm -- (Kyoto: Kyoto University Press)\par
Kato, S., Okazaki, A.-T., \& Oktariani, F. 2011, 63, 365 \par
%Lai, D. \& Tsang, D. 2009, MNRAS, 393, 979 \par
Lamb, H. 1924, Hydrodynamics (Cambridge; Cambrigde University Press) p.335\par
Lubow, S.H. 1991, ApJ, 381, 259\par
Lubow, S.H. 1992, ApJ, 401, 317\par
%Lynden-Bell, D. and Ostriker, J.P. 1967, MNRAS, 136, 293  \par
Martin, R.G., Pringle, J.E., Tout, C.A., \& Lubow, S.H. 2011, MNRAS, 416, 2827\par
Meyer-Vernet, N. \& Sicardy, B. 1987, Icarus, 69, 157 \par
Moritani, Y., Nogami, D., Okazaki, A.T., Imada, A., Kambe, E., Honda, S., Hashimoto, O., Mizoguchi, S., Kanda, Y.,     
      Sadakakane, K., \& Ichikawa, K. 2013, PASJ, 65, 83 \par 
Morse, P. M., \& Feshbach, H. 1953, Methods of Theoretical Physics (McGraw Hill Comp, New York), chapter 9.3\par 
Negueruela, I., \& Okazaki, A.T. 2001, A\&A, 369,108\par
Okazaki, A.T. 1991, PASJ 43, 750 \par
Okazaki, A.T., Kato, S., \& Fukue, J. 1987, PASJ, 39, 457 \par
Okazaki, A.T., \& Negueruela, I. 2001, A\& A 2001, 377, 161 \par
Okazaki, A.T., Hayasaki, K., \& Moritani, Y. 2013,  PASJ, 65, 41\par
Osaki, Y. 1985, A\&A, 144, 369\par
%Oktariani, F., Okazaki, A.-T., \& Kato, S. 2010, PASJ, 62, xxx \par
Ortega-Rodr\'{i}guez, M., Silbergleit, A.S., \& Wagoner, R.V. 2008, Geophys. Astrophys. Fluid Dynamics, 102,75\par
Reig, P. 2011, Ap\&SS, 332, 1 \par 
Wood, K., Bjorknan, K.,S., \& Bjorkman, J.E. 1997, ApJ, 477, 926\par     

\end{document}